\documentclass[graybox]{svmult}

\usepackage[varg]{txfonts}

\usepackage{helvet}         % selects Helvetica as sans-serif font
\usepackage{courier}        % selects Courier as typewriter font
\usepackage{type1cm}        % activate if the above 3 fonts are
\usepackage{makeidx}         % allows index generation
\usepackage{graphicx}        % standard LaTeX graphics tool
\usepackage{multicol}        % used for the two-column index
\usepackage[bottom]{footmisc}% places footnotes at page bottom

\usepackage{amssymb}
\makeindex             % used for the subject index
\begin{document}

\title*{Reconciling a significant hierarchical assembly of massive early-type galaxies at z$\lesssim$1 with mass downsizing}
\titlerunning{Mass downsizing and the significant hierarchical assembly of mETGs since z$\sim$1} 

\author{M.Carmen Eliche-Moral, Mercedes Prieto, Jes\'us Gallego, \& Jaime Zamorano}
\authorrunning{Eliche-Moral et al.}

\institute{M.\,C.\,Eliche-Moral, J.\,Gallego, \& J.\,Zamorano \at Departamento de Astrof\'{\i}sica, Universidad Complutense de Madrid, E-28040, Madrid, Spain\\\email{mceliche@fis.ucm.es}
\and M.\,Prieto \at Instituto de Astrof\'{\i}sica de Canarias, C/ V\'{\i}a L\'actea, E-38200, La Laguna, Tenerife, Spain}

\maketitle

\vskip-1.2truein

\abstract{
Hierarchical models predict that massive early-type galaxies (mETGs) are the latest systems to be in place into the cosmic scenario (below z$\sim$0.5), conflicting with the observational phenomenon of galaxy mass downsizing, which poses that the most massive galaxies have been in place earlier that their lower-mass counterparts (since z$\sim$0.7). We have developed a semi-analytical model to test the feasibility of the major-merger origin hypothesis for mETGs, just accounting for the effects on galaxy evolution of the major mergers strictly reported by observations. The most striking model prediction is that very few present-day mETGs have been really in place since z$\sim$1, because $\sim$90\% of the mETGs existing at z$\sim$1 are going to be involved in a major merger between z$\sim$1 and the present. Accounting for this, the model derives an assembly redshift for mETGs in good agreement with hierarchical expectations, reproducing observational mass downsizing trends at the same time.}

\vspace{-0.3cm}

\section{Assembly of mETGs in groups and field}
\label{sec:assembly}

Recent observations disagree with the late major-merger origin that hierarchical models predict for mETGs (\cite{2006MNRAS.366..499D}). Massive galaxies ($\log(\mathcal{M}_*/\mathcal{M}_\odot)\gtrsim$11, mostly E-S0a's) seem to have been already in place at z$\sim$0.7, earlier than their less-massive counterparts (this phenomenon is known as mass downsizing, see \cite{2008ApJ...687...50P}). Moreover, the S0 fraction in clusters increases with time at the expense of the spirals, being the fraction of ellipticals nearly constant (\cite{2009ApJ...697L.137P}). This has been interpreted as a sign of the poor role played by major mergers in the evolution of cluster ellipticals (contrary to the hierarchical picture). However, more than $\sim$1/2 of present-day mETGs do not reside in clusters, but in groups (\cite{2006ApJS..167....1B}), where mergers and tidal interactions determine the galaxy evolution (\cite{2008ApJ...688L...5K}). So, the question is still open: are major mergers really the main drivers of the assembly of most mETGs? We have used semi-analytical models to test this hypothesis, studying how the present-day mETGs would have evolved backwards-in-time assuming that they derive exclusively from the (wet/mixed/dry) major mergers strictly reported by observations. The model predictions apply to field and group mETGs, as the local luminosity functions (LFs) and major merger fractions used here trace basically these environments (\cite{2010A&A...519A..55E,2010arXiv1003.0686E}).

\vspace{-1cm}

\section{Reconciling mass downsizing and hierarchical scenarios}
\label{sec:reconciling}

\begin{figure}[t]
\sidecaption[t]
\includegraphics[scale=.95]{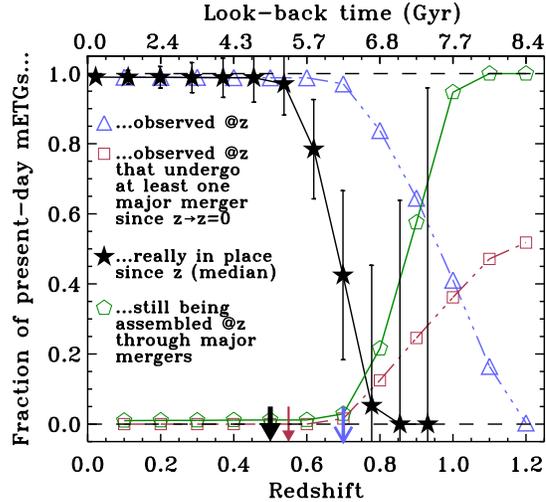}
\caption{Model predictions on the fraction of present-day mETGs observed, really in place, and still being assembled at each redshift \emph{z}. \emph{Arrows}: Assembly redshift of mETGs according to observations (\emph{open}, \cite{2008ApJ...687...50P}), to the model (\emph{thick solid}), and average $z$ between the assembly redshifts of 50\% and 80\% of their stellar content according to the models by \cite{2006MNRAS.366..499D} (\emph{thin solid}). Our model reproduces the observed buildup of mETGs since z$\sim$1 (triangles), predicting an assembly redshift for mETGs in good agreement with hierarchical models (\cite{2006MNRAS.366..499D}) at the same time.}
\label{fig:fig1}       
\end{figure}

The model can reproduce the observed evolution of the galaxy LFs at z$\lesssim$1, simultaneously for different rest-frame bands and selection criteria (based on color or morphology). The model predicts that $\sim$90\% of the mETGs existing at $z\sim 1$ are not the passively-evolved high-z counterparts of present mETGs (as usually interpreted in many studies), but their gas-poor progenitors instead. This implies that very few present-day mETGs have been really in place since z$\sim$1 ($\lesssim$5\%), contrary to the $\sim$50\% reported by traditional interpretations of observations. Accounting for this, the model is capable of deriving a final assembly redshift of most mETGs in good agreement with hierarchical models (z$\sim$0.5, neglecting minor mergers), reproducing global mass downsizing trends at the same time (see Fig.\,\ref{fig:fig1}, adapted from \cite{2010arXiv1003.0686E}). 

\vspace{-0.3cm}

\begin{acknowledgement}
Supported by the Spanish Ministry of Science and Innovation (MICINN) under project AYA2009-10368 and by the Madrid Regional Government through the AstroMadrid Project (CAM S2009/ESP-1496). Funded by the Spanish MICINN under the Consolider-Ingenio 2010 Program grant CSD2006-00070: "First Science with the GTC".
\end{acknowledgement}
\end{document}